# Time preference, wealth and utility inequality: A microeconomic interaction and dynamic macroeconomic model connection approach

Takeshi Kato*

Hitachi Kyoto University Laboratory, Open Innovation Institute, Kyoto University

Address: Yoshidahonmachi, Sakyo-ku, Kyoto-shi, Kyoto 606-8501, Japan

Tel: + 81-75-753-9716

E-mail: kato.takeshi.3u@kyoto-u.ac.jp

ORCiD: 0000-0002-6744-8606

## Abstract

Based on interactions between individuals and others and references to social norms, this study reveals the impact of heterogeneity in time preference on wealth distribution and inequality. We present a novel approach that connects the interactions between microeconomic agents that generate heterogeneity to the dynamic equations for capital and consumption in macroeconomic models. Using this approach, we estimate the impact of changes in the discount rate due to microeconomic interactions on capital, consumption and utility and the degree of inequality. The results show that intercomparisons with others regarding consumption significantly affect capital, i.e. wealth inequality. Furthermore, the impact on utility is never small and social norms can reduce this impact. Our supporting evidence shows that the quantitative results of inequality calculations correspond to survey data from cohort and cross-cultural studies. This study's micro–macro connection approach can be deployed to connect microeconomic interactions, such as exchange, interest and debt, redistribution, mutual aid and time preference, to dynamic macroeconomic models.

# INTRODUCTION

Wealth inequality has become a significant social problem worldwide. According to the World Inequality Report 2022 (Chancel *et al.*, 2022), the top 1% of the wealthiest people account for 38% of the world's wealth, while the World Economic Forum (2017) found that only eight men have the same wealth as the poorest 3.6 billion people. The global Gini index has reached 0.7 (Chancel *et al.*, 2022), well above the warning level of 0.4 for social unrest (United Nations Human Settlements Programme). Solving inequality is important because social unrest generates a vicious cycle that reduces productivity, increases inequality and fuels social unrest (Sedik & Xu, 2020).

According to the macroeconomics review literature on heterogeneity (Heathcote *et al.*, 2009), efforts to focus on microeconomic heterogeneity in macroeconomics have three main directions: individual endogenous heterogeneity and exogenous shocks, individual insurance against exogenous risks and the interaction between idiosyncratic risks and aggregate dynamics. Concerning sources of heterogeneity, efforts have been made regarding endogenous capabilities and preferences as initial conditions for individuals and transitory and persistent shocks as exogenous conditions.

The importance of heterogeneity in individuals' time preferences for wealth inequality is well known (Frederick *et al.*, 2002). Individual wealth accumulation and consumption are strongly influenced by how future values are discounted and heterogeneity in time preferences leads to wealth inequality. Considerable research has investigated the relationship between time preference and inequality. The representative literature includes a model in which heterogeneity in discount rates leads to wealth concentration (Becker, 1980), an empirical study of the relationship between heterogeneity in time preference and inequality (Lawrance, 1991) and a model to explain inequality from the distribution of discount rates (Krusell & Smith, 1998). Additional studies include a model to estimate heterogeneity in time preferences from the distribution of inequality (Hendricks, 2007), a model to account for inequality by adding consumer choice and human capital formation to heterogeneity in time preferences (Suen, 2014) and an empirical study of the relationship between differences in saving behaviour and inequality based on the distribution of time preferences (Epper *et al.*, 2020).

Two main approaches exist to studying utility functions, including time preference: the constant and endogenous discount rates (Shi & Epstein, 1993). The constant rate is a model that expresses instantaneous utility as consumption versus reference stock under a constant discount rate (Ryder & Heal, 1973; Carroll *et al.*, 1997; Strulik, 2008). In contrast, the endogenous discount rate adds the effect of endogenous habit formation to the discount rate for instantaneous utility (Uzawa, 1968; Epstein, 1987; Obstfeld, 1990; Chang *et al.*, 2011; Zhang, 2012). In dealing with heterogeneity in time preference, the model that deals with endogenous changes in discount rates is more suitable than the model that deals with constant discount rates.

Reviewing the literature on the measurement of time preference (Frederick *et al.*, 2002; Cohen *et al.*, 2020) indicates three main measurement models: the exponential, hyperbolic and quasi-hyperbolic forms. The classical exponential form is a function with the discount rate as the base and time as the exponent (Samuelson, 1937). The hyperbolic form is a function with the linear function of the term multiplied by the discount rate and time as the base and the inverse of the discount rate as the exponent (Prelec & Loewenstein, 1991). The quasi-hyperbolic form is a function of the exponential form multiplied by the present bias factor (Laibson, 1997). Wang *et al.* (2016) measured time preference in 53 countries based on quasi-hyperbolic forms, which Rieger *et al.* (2021) combined with the sociological and social psychology survey literature to obtain universal time preference in 117 countries.

Previous studies on time preference have identified the introduction of various heterogeneities (Sedik & Xu, 2020), reflecting the interaction of multiple psychological motivations (Frederick *et al.*, 2002) and considering psychological factors such as trust, confusion and context (Cohen *et al.*, 2020), as future challenges. Concerning these issues, social psychology has traditionally studied the close relationship between self and others (Markus & Wurf, 1987), while cognitive science has investigated the social brain hypothesis that social groups influence the human brain (Dunbar, 2003). The comparative institutional analysis of microeconomics posits that shared expectations constrain individual strategies and behaviours as a summary expression of the group's equilibrium state (Aoki, 2001). Additionally, the theory of social norms related to public economics states that social norms, including customs, manners and rules, broadly originate from mutual expectations and approval with others based on sympathetic human tendencies (Heath, 2008). Therefore, it is essential to consider interactions with others and social groups as psychological motivations and factors related to time preferences; however, while conventional economic studies on time preference have examined the effects on individuals' endogenous habit formation and preference changes for exogenous shocks, no studies focus on others' interactions and social context.

The agent-based model (ABM) is an approach that introduces microeconomic interactions into the macroeconomy. The related literature (Haldane & Turrell, 2018) states that macroeconomic ABM attempts to understand macro-level behaviour by having rule-based interactions among agents (Turrell, 2016). Representative studies show phase transitions between low and high-unemployment economies by setting firm production and price rules (Gualdi *et al.*, 2015). Additional studies have shown that policy indicators such as gross domestic product, growth and unemployment rates by modelling the behaviour of government, bank, firm and household agents (Caiani *et al.*, 2016) and that business cycles are due to the interaction of agents' decision-making with cognitive restrictions (De Grauwe, 2010). Note that agents' behaviour in macroeconomic ABM is rule-based and autonomous dynamic behaviour is discarded.

Econophysics is an approach that uses ABM and focuses specifically on wealth distribution and inequality. The literature in this realm (Chatterjee & Chakrabarti, 2007; Chakrabarti AS & BK, 2010) shows that asset exchanges among many agents are modelled based on physical analogies to simulate various wealth distributions. Representative studies have shown that the gamma distribution is caused by the stochastic process of the exchange of assets other than savings (Angle, 1986) and that the delta distribution uses the analogy of the kinetic energy exchange of gas particles (Chakraborti, 2002). Further research has shown that changes in tax rates and insurance can cause exponential and gamma distributions to interchange (Guala, 2009; Chakrabarti AS & BK, 2010). No macroeconomic models are incorporated into these econophysical agents.

This study aims to simulate the distribution and inequality of wealth and utility by setting up a new dynamic model concerning both econophysics and macroeconomic

ABM. We also investigate the heterogeneity of time preference, i.e. discount rate, based on the interaction between individuals and others and the social context and norms. Econophysics has a physical analogy-based agent that differs from reality and the macroeconomic ABM has a behavioural rule-based agent. In contrast, this study takes the approach of connecting microeconomic interactions with dynamic macroeconomic agents to incorporate the dynamic behaviour of autonomous agents. Unlike previous studies that calculate wealth inequality from the distribution of pre-given time preferences (Suen, 2014), examine the relationship between measured time preference distributions and inequality (Epper *et al.*, 2020) or change time preferences through endogenous habit formation (Zhang, 2012), this study takes a unique approach. We examine the distribution and inequality of wealth and utility generated by the distribution of time preferences generated from initial uniform time preferences through interactions with others and social norms. Our approach is limited in that it assumes an impact on time preference by its interactions and social norms, but as with Zhang's habit formation, the actual survey researches suport the findings obtained.

The remainder of this paper is organized as follows. The Methods section first refers to the Ramsey–Cass–Koopmans model (Ramsey, 1928; Cass, 1965; Koopmans, 1965), one of the basic models of dynamic macroeconomics. On this basis, we show how to calculate the dynamic saddle path and the utility for discount rate changes concerning capital and individual agents' consumption. Following an econophysical approach, we then formulate changes in the discount rates based on intercomparisons of capital or consumption among agents and references to social norms. Through repeated interactions and references, the Results section simulates changes in the distribution of discount rates from initial uniform values and corresponding changes in the distribution of wealth and utility. This section also presents the calculation results of coefficients of variation and Gini indices that represent inequality among agents. The Discussion section considers the impact of intercomparisons and social norms on inequality, along with the interpretation of the results and mentions the potential of developing an approach that connects microeconomic interactions other than time preference, such as exchange, debt, redistribution and mutual aid, to dynamic macroeconomics. Finally, the Conclusion section presents the main inferences and addresses future issues.

## METHODS

**Dynamic macroeconomic model**

This study refers to the dynamic equations for capital and consumption in the Ramsey–Cass–Koopmans model as the basic dynamic macroeconomic model (Romer, 2019: 59–60, 95; Nakata, 2011: 45). Suppose capital per individual (household) is $k$, consumption is $c$, the production function is $f(k)$, the capital depletion rate is $\delta$, the labour growth rate is $\lambda$, the knowledge growth rate is $\kappa$, the discount rate of time preference is $\rho$ and the relative risk aversion coefficient is $\theta$. In this case, the dynamic equations for capital $k$ and consumption $c$ concerning time $t$ are expressed as equations (1) and (2).

$$\dot{k}(t) = f(k(t)) - c(t) - (\delta + \lambda + \kappa)k(t). \tag{1}$$

$$\frac{\dot{c}(t)}{c(t)} = \frac{f'(k(t)) - \delta - \rho - \theta\kappa}{\theta}. \tag{2}$$

This study uses the Cobb–Douglas function $f(k(t)) = k(t)^\alpha$ as the production function (Romer, 2019: 13; Nakata, 2011: 121) and setting $\lambda = \kappa = 0$ because the labour and knowledge growth rates are not covered. Then, the dynamic equations can be rewritten as equations (3) and (4).

$$\dot{k}(t) = k(t)^\alpha - c(t) - \delta k(t). \tag{3}$$

$$\dot{c}(t) = c(t)\frac{\alpha k(t)^{\alpha-1} - \delta - \rho}{\theta}. \tag{4}$$

These dynamic equations are known to converge at the saddle point (Romer, 2019: 62–65; Nakata, 2011: 122). If we solve the equations with the left-hand side of (3) and (4) set to zero, the capital $k^*$ and consumption $c^*$ at the saddle point can be expressed as equations (5) and (6).

$$k^* = \left(\frac{\delta + \rho}{\alpha}\right)^{\frac{1}{\alpha-1}}. \tag{5}$$

$$c^* = k^{*\alpha} - \delta k^*. \tag{6}$$

We now assume the case where the discount rate changes (Romer, 2019: 67–69). Let

$\rho_{OLD}$ be the discount rate before the change and $\rho_{NEW}$ be the discount rate after the change. Then, using equations (5) and (6), the saddle points $k^*_{OLD}$ and $c^*_{OLD}$ before the change are expressed as equations (7) and (8); the saddle points $k^*_{NEW}$ and $c^*_{NEW}$ are expressed as equations (9) and (10) after the change.

$$k^*_{OLD} = \left(\frac{\delta + \rho_{OLD}}{\alpha}\right)^{\frac{1}{\alpha-1}}. \tag{7}$$

$$c^*_{OLD} = k^{*\alpha}_{OLD} - \delta k^*_{OLD}. \tag{8}$$

$$k^*_{NEW} = \left(\frac{\delta + \rho_{NEW}}{\alpha}\right)^{\frac{1}{\alpha-1}}. \tag{9}$$

$$c^*_{NEW} = k^{*\alpha}_{NEW} - \delta k^*_{NEW}. \tag{10}$$

When the discount rate changes from $\rho_{OLD}$ to $\rho_{NEW}$, capital $k^*_{OLD}$ initially maintains the same value because it cannot change discontinuously and consumption $c^*_{OLD}$ jumps discontinuously (Romer, 2019: 67–69). At this time, capital $k_A$ and consumption $c_A$ are expressed as equations (11)–(14) (Romer, 2019: 69–71).

$$k_A = k^*_{OLD}. \tag{11}$$

$$c_A = c^*_{NEW} + \frac{f''(k^*_{NEW})c^*_{NEW}}{\theta\mu}(k^*_{OLD} - k^*_{NEW}). \tag{12}$$

$$f''(k^*_{NEW}) = \alpha(\alpha - 1)k^{*\alpha-2}_{NEW}. \tag{13}$$

$$\mu = \frac{\rho_{NEW} - \sqrt{\rho_{NEW}^2 - 4f''(k^*_{NEW})c^*_{NEW}/\theta}}{2}. \tag{14}$$

After the change, capital $k_A$ and consumption $c_A$ converge along an adjustment path toward the new saddle points $k^*_{NEW}$ and $c^*_{NEW}$ (Romer, 2019: 69–71). Along this adjustment path, the changes in capital $k_{ADJ}(t)$ and consumption $c_{ADJ}(t)$ are represented by the approximate equations (15) and (16).

$$k_{ADJ}(t) = k^*_{NEW} + e^{\mu t}(k_A - k^*_{NEW}). \tag{15}$$

$$c_{ADJ}(t) = c^*_{NEW} + e^{\mu t}(c_A - c^*_{NEW}). \tag{16}$$

**Utility model**

We next refer to the Ramsey–Cass–Koopmans infinite-horizon model as the basic utility model (Romer, 2019: 52–55; Nakata, 2011: 40–44). The infinite-horizon utility function $U$ is expressed as equation (17). Here, the discount rate is $\rho$, the instantaneous utility is $u(c(t))$ and the instantaneous utility, $u(c(t))$, is $c(t)^{1-\theta}/(1-\theta)$ with constant relative risk aversion.

$$U = \int_{t=0}^{\infty} e^{-\rho t} u(c(t)) dt = \int_{t=0}^{\infty} e^{-\rho t} \frac{c(t)^{1-\theta}}{1-\theta} dt. \qquad (17)$$

For a measure of time preference based on the quasi-hyperbolic form, the utility function $U$ is expressed as equation (18). Here, patience is $u_0$, the current bias is $\beta_p$ and the discount rate is $\phi$ (Wang et al., 2016). Rewriting $\phi^t$ as an exponential function with $e$ as the base, $-\ln(1/\phi)t$ as the exponent and setting $u_0 = 0$ and $\beta_p = 1$ for simplicity, equation (18) coincides with (17).

$$\begin{aligned} U &= u_0 + \int_{t=0}^{\infty} \beta_p \phi^t u(c(t)) dt \\ &= u_0 + \beta_p \int_{t=0}^{\infty} e^{-\ln\left(\frac{1}{\phi}\right)t} u(c(t)) dt. \end{aligned} \qquad (18)$$

We now assume that the discount rate changes $n$ times in the middle of an infinite horizon. Let the time of change be $t_1, t_2, \cdots, t_n$ and the discount rate be $\rho_0, \rho_1, \cdots, \rho_n$. Then, the utility function, $U$, is expressed as equation (19).

$$\begin{aligned} U = &\int_0^{t_1} e^{-\rho_0 t} \frac{c_0(t)^{1-\theta}}{1-\theta} dt + \int_{t_1}^{t_2} e^{-\rho_1 t} \frac{c_1(t)^{1-\theta}}{1-\theta} dt + \cdots \\ &+ \int_{t_n}^{\infty} e^{-\rho_n t} \frac{c_n(t)^{1-\theta}}{1-\theta} dt \end{aligned} \qquad (19)$$

**The interaction model of time preference**

We refer to the econophysics approach in modelling interactions with others and referencing social norms regarding time preferences. In econophysics (Chatterjee & Chakrabarti, 2007; Chakrabarti, AS & BK, 2010), two agents are randomly selected from a large number of agents, the exchange of assets between them is modelled and the repeated selection of two agents is used to simulate various wealth distributions. Thus, we model the change in time preference due to the interaction between agents $i$ and $j$. Let the discount rates of both agents at time $t$ be $\rho_{OLD_i}(t)$ and $\rho_{OLD_j}(t)$; both change to $\rho_{NEW_i}(t)$ and $\rho_{NEW_j}(t)$ due to their interaction in equations (20) and (21), respectively.

$$\rho_{NEW_i}(t) = \rho_{OLD_i}(t)\left(1 - \varepsilon_k \frac{\beta_k k_j(t) - k_i(t)}{k_i(t)} + \varepsilon_c \frac{\beta_c c_j(t) - c_i(t)}{c_i(t)} + \varepsilon_\rho \frac{\rho_{NORM} - \rho_i(t)}{\rho_i(t)}\right). \quad (20)$$

$$\rho_{NEW_j}(t) = \rho_{OLD_j}(t)\left(1 - \varepsilon_k \frac{\beta_k k_i(t) - k_j(t)}{k_j(t)} + \varepsilon_c \frac{\beta_c c_i(t) - c_j(t)}{c_j(t)} + \varepsilon_\rho \frac{\rho_{NORM} - \rho_j(t)}{\rho_j(t)}\right). \quad (21)$$

The second and third terms in the brackets on the right-hand side of equations (20) and (21) represent the interaction based on the intercomparison of capital $k_i(t)$ and $k_j(t)$ and consumption $c_i(t)$ and $c_j(t)$ (t), respectively, because the time preference is considered to be influenced by both capital (future consumption) and current consumption (Romer, 2019: 370–371). $\beta_k \geq 1$ and $\beta_c \geq 1$ are parameters that represent the so-called 'the grass is greener on the other side' psychological bias. The existence of this psychological bias is well known in neuroscience (Boorman *et al.*, 2009; Noritake *et al.*, 2018); however, its effects on capital and consumption remain unexamined; therefore, we formulate them here as in equations (20) and (21). Here, $0 \leq \varepsilon_k \leq 1$ and $0 \leq \varepsilon_c \leq 1$ are the parameters for the strength of the interaction. Equation (20) explains that if $\beta_k k_j(t) > k_i(t)$ when others' capital appears larger, the discount rate $\rho_{NEW_i}(t)$ becomes smaller than $\rho_{OLD_i}(t)$ (reducing consumption). Furthermore, if $\beta_c c_j(t) > c_i(t)$ when others' consumption appears larger, $\rho_{NEW_i}(t)$ becomes larger than $\rho_{OLD_i}(t)$ (increasing consumption).

The fourth term represents the reference to the social normative discount rate, $\rho_{NORM}$, where $0 \leq \varepsilon_\rho \leq 1$ is a parameter indicating the strength of the reference. Social norms include secular morality, Buddhist teachings and Islamic law. If $\varepsilon_k = \varepsilon_c = 0$ and $\varepsilon_\rho = 1$, then $\rho_{NEW_i}(t) = \rho_{NEW_j}(t) = \rho_{NORM}$. Equation (20) explains that if

$\rho_{OLD_i}(t) > \rho_{NORM}$ when the current discount rate is larger than the normative discount rate, then $\rho_{NEW_i}(t)$ becomes smaller than $\rho_{OLD_i}(t)$ (reducing consumption). In other words, the normative discount rate $\rho_{NORM}$ works in a suppressive manner; however, if $\rho_{NORM}$ is large (when the whole society is ephemeral), $\rho_{NEW_i}(t)$ will become larger (increasing consumption).

Figure 1 shows the computational flow for time preference interaction. First, in Step I, the number of agents ($N$), the time increment ($\Delta t$) of the computation, the period ($t_p$) for the interaction and the maximum time ($t_{max}$) for the computation are set as initial settings. Furthermore, the initial discount rate ($\rho_0$), the initial utility ($U_0$), the exponent ($\alpha$) of the production function, the capital depletion rate ($\delta$) and the coefficient of relative risk aversion ($\theta$) are set as initial settings for all agents. Although not shown in Figure 1, parameters $\beta_k$, $\beta_c$, $\varepsilon_k$, $\varepsilon_c$ and $\varepsilon_\rho$ in equations (20) and (21) are also set together. In Step II, we use the initial values and equations (5) and (6) and calculate the capital ($k_0^*$) and consumption ($c_0^*$) for all agents at time $t = 0$. In Step III, we advance time $t$ to $t + \Delta t$. In Step IV, we calculate the surplus $Mod(t/t_p)$ of time $t$ and divided by $t_p$. If it is 0, we return to Step V; if the value is not 0, we proceed to Step IX.

In Step V, we randomly select two agents $i$ and $j$ ($i \neq j, i, j = 1, 2, \cdots, N$) among $N$ agents. In Step VI, using equations (20) and (21), we change the discount rates $\rho_{OLD_i}(t)$ and $\rho_{OLD_j}(t)$ of agents $i$ and $j$ to $\rho_{NEW_i}(t)$ and $\rho_{NEW_j}(t)$, respectively. In Step VII, using equations (11) and (12), we change the capital $k_{OLD_i}(t)$ and consumption $c_{OLD_i}(t)$ for agent $i$ and $k_{OLD_j}(t)$ and $c_{OLD_j}(t)$ for $j$ to $k_{A_i}(t)$, $c_{A_i}(t)$, $k_{A_j}(t)$ and $c_{A_j}(t)$, respectively. In Step VIII, using equations (15) and (16), we set the adjustment paths $k_{ADJ_i}(t)$, $c_{ADJ_i}(t)$, $k_{ADJ_j}(t)$ and $c_{ADJ_j}(t)$ toward the new saddle points $k_{NEW_i}(t)$, $c_{NEW_i}(t)$, $k_{NEW_j}(t)$ and $c_{NEW_j}(t)$, which are calculated using equations (9) and (10). After that, as time $t$ advances, $k_i(t)$, $c_i(t)$, $k_j(t)$ and $c_j(t)$ proceed along the adjustment path.

In Step IX, which follows from Step IV or VIII, we calculate the increment of utility $U_i$ corresponding to time increment $\Delta t$ for each agent $i$ ($i = 1, 2, \cdots, N$) based on equations (17) or (19); we use the discount rate $\rho_i(t)$ and instantaneous utility

$c_i(t)^{1-\theta}/(1-\theta)$ at time $t$. In Step X, if time $t$ has reached $t_{max}$, we proceed to Step XI; otherwise, we return to Step III. In Step XI, we calculate the utility from time $t_{max}$ to infinite horizon $\infty$ using the discount rate $\rho_i(t_{max})$ at time $t_{max}$, added to utility $U_i$ and the process ends. If the discount rate is changed again by returning to Step III and proceeding from IV to VI, the new saddle point is recalculated and $k_i(t)$, $c_i(t)$, $k_j(t)$ and $c_j(t)$ proceed along the new adjustment path.

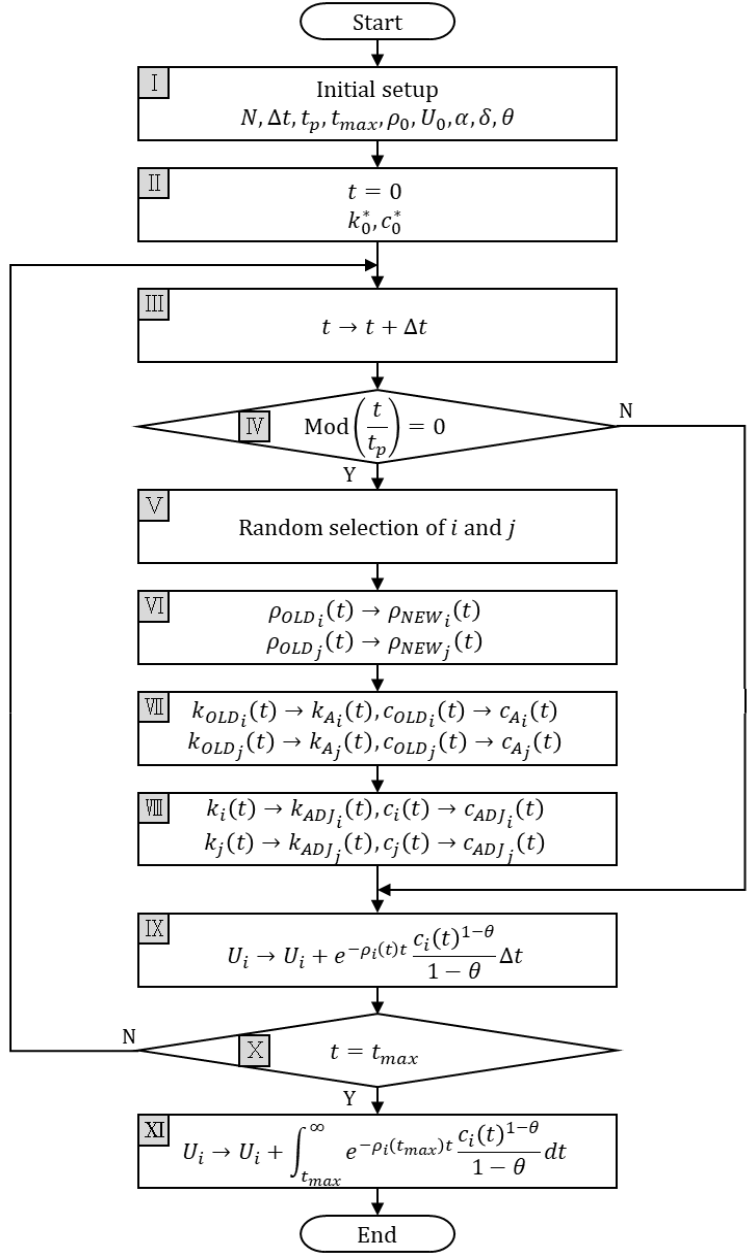

**Figure 1:** Computational flow for time preference interaction

# RESULTS

First, we initialize the common parameters to be used in the calculations. The number of agents is set to $N = 1,000$ (sufficient to compute the wealth distribution and Gini index). The time increment is $\Delta t = 1/(24 \times 365)$ (hour), the interaction period is $t_p = 1/365$ (day) and the maximum computation time is $t_{max} = 10$ (years). These values are relative; therefore, they could be read as 10 hours, 10 days and 100 years, respectively. Over 10 years, 7,300 people ($10 \times 365 \times 2$) would change their behaviour, each receiving an average of 0.73 interactions per year.

The initial discount rate is set to $\rho_0 = \ln(1/\phi) = 0.223$ using the value $\phi \sim 0.8$ from the literature (Wang *et al.*, 2016) and equation (18). The normative discount rate is set to $\rho_{NORM} = 0.2$, which is smaller than the initial value and the initial utility is set to $U_0 = 0$. We set the exponent $\alpha$ and capital depletion rate $\delta$ of the production function to $\alpha = 0.5$ and $\delta = 0.1$ respectively, based on the values used in the literature (Romer, 2019: 27, 72; Nakata, 2011: 9, 90, 94; Hasumi, 2020: 25–26, 71). For the coefficient of relative risk aversion ($\theta$), a wide variety of values have been reported depending on the measurement method and subject (Hartog *et al.*, 2002; Itou, 2008; Elminejada *et al.*, 2022); however, here it is set to $\theta = 0.5$. Initial capital and consumption are calculated as $k_0 = 2.39$ and $c_0 = 1.31$ using equations (5) and (6), respectively.

We used the following 12 calculations for the discount rate, the others' interaction parameters ($\varepsilon_k$, $\varepsilon_c$) and the social norm reference parameter ($\varepsilon_\rho$): (i) three cases of $\varepsilon_k = 0.1$, $0.2$ or $0.3$ ($\varepsilon_c = \varepsilon_\rho = 0$); (ii) three cases of $\varepsilon_c = 0.1$, $0.2$ or $0.3$ ($\varepsilon_k = \varepsilon_\rho = 0$); (iii) three cases where the setting in (i) changes to $\varepsilon_\rho = 0.1$; and (iv) three cases where the setting in (ii) is changed to $\varepsilon_\rho = 0.1$. Primarily, the strength of the interaction is expressed in terms of $\varepsilon_k$ and $\varepsilon_c$ and the constant $\beta_k = \beta_c = 1.1$ are set for the psychological bias parameter.

Figure 2 shows the calculated distributions of discount rate, capital, consumption and utility for capital interactions $\varepsilon_k = 0.1$, $\varepsilon_c = \varepsilon_\rho = 0$ and $\beta_k = 1.1$. Figure 2A plots the discount rate $\rho(t_{max})$ at time $t_{max} = 10$ for agent number #, while Figure 2B shows the frequency distribution of discount rate $\rho(t_{max})$. Similarly, Figures 2C and 2D show the distribution of capital $k(t_{max})$, while 2E and 2F show the consumption $c(t_{max})$ and 2G and 2H show the utility $U$ over an infinite horizon. Figure 2B shows that comparing capital with others, the mean of discount rate $\rho(t_{max})$ decreases from the initial $\rho_0 = 0.223$ to 0.207 (future-oriented). Accordingly, Figures 2D and 2F show that the mean of capital $k(t_{max})$ and consumption $c(t_{max})$ increase from initial $k_0 = 2.39$ and $c_0 = 1.31$ to 2.65 and 1.36, respectively. The skewness and kurtosis of the distribution of $\rho(t_{max})$ are, respectively, 0.44 and 3.30, $-0.33$ and 3.11 for $k(t_{max})$, $-0.40$ and 3.23 for $c(t_{max})$ and $-1.27$ and 5.96 for $U$. Except for $U$, the others are generally normal distributions.

Figure 3 shows the temporal changes in discount rate, capital, consumption and utility for $\varepsilon_k = 0.1$ and $\varepsilon_c = \varepsilon_\rho = 0$, as in Figure 2. Figure 3A plots the discount rate $\rho$ for time $t$ concerning agent #1 and Figure 3B is an enlarged part of 3A. Similarly, Figures 3C and 3D show the capital $k$ and 3E and 3F show the consumption $c$. Figures 3G and 3H are the values of the interval integral of utility $U$ from time 0 to $t$. Note that the value of utility $U$ converges to a constant value at infinite horizon $\infty$. Figure 3A shows that agent #1 received nine interactions. In Figure 3B, the discount rate $\rho$ decreases around time $t = 7.0$ and accordingly, Figure 3F shows that the consumption $c$ decreases discontinuously. Thereafter, Figures 3D and 3F indicate that capital $k$ and consumption $c$ follow an adjustment path and converge to a new saddle point. At the new saddle point, capital $k$ and consumption $c$ are higher than before the decrease in discount rate $\rho$ (Romer, 2019: 67–69). In other words, comparing capital with others suppresses the agent's consumption once; however, it leads to a higher level of economic behaviour in the future. Figure 3H shows a slight inflexion in the curve of utility $U$ due to the change in the discount rate $\rho$.

Figure 4 shows the calculated distributions of discount rate, capital, consumption and utility for consumption interactions $\varepsilon_c = 0.1$, $\varepsilon_k = \varepsilon_\rho = 0$ and $\beta_c = 1.1$. Figure 4A plots the discount rate $\rho(t_{max})$ for agent number # and Figure 4B shows the frequency distribution of discount rate $\rho(t_{max})$. Similarly, Figures 4C and 4D show the capital $k(t_{max})$, 4E and 4F show the consumption $c(t_{max})$ and 4G and 4H show the utility $U$. By comparison of consumption with others, in Figure 4B, the mean of discount rate $\rho(t_{max})$ increases from the initial $\rho_0 = 0.223$ to 0.240 (future despised). Accordingly, Figures 4D and 4F show that the mean of capital $k(t_{max})$ and consumption $c(t_{max})$ decrease from initial $k_0 = 2.39$ and $c_0 = 1.31$ to 2.17 and 1.25, respectively. This finding is opposite to the trend in Figure 2. The skewness and kurtosis of the distribution of $\rho(t_{max})$ are 1.00 and 4.59, respectively, $-0.73$ and 3.72 for $k(t_{max})$, $-0.85$ and 4.08 for $c(t_{max})$ and $-1.42$ and 5.99 for $U$. These values are all larger than that in Figure 2, indicating that consumption interaction $\varepsilon_c$ distorts the shape of the distribution more than capital interaction $\varepsilon_k$.

Figure 5 shows the temporal changes in discount rate, capital, consumption and utility for $\varepsilon_c = 0.1$ and $\varepsilon_k = \varepsilon_\rho = 0$, as in Figure 4. Figure 5A plots the discount rate $\rho$ for time $t$ concerning agent #1 and Figure 5B is an enlarged part of 5A. Similarly, Figures 5C and 5D show the capital $k$ and 5E and 5F show the consumption $c$. Figures 5G and 5H are the values of the interval integral of utility $U$ from time 0 to $t$. Figure 5A shows that agent #1 received four interactions. In Figure 5B, the discount rate $\rho$ increases around time $t = 7.8$. Accordingly, Figure 5F shows that the consumption $c$ once increases discontinuously. After that, as in Figures 5D and 5F, capital $k$ and consumption $c$ follow an adjustment path and converge to a new saddle point. At the new saddle point, capital $k$ and consumption $c$ are lower than before the increase in discount rate $\rho$. This finding is opposite to the trend in Figure 3. The comparison of consumption with others induces the agent's consumption once; however, it leads to a lower level of economic behaviour in the future. Figure 5H shows a slight inflexion in the curve of utility $U$ due to the change in the discount rate $\rho$.

Figures 6A–6D show the results of calculating the mean and coefficient of variation (CV) of the discount rate $\rho(t_{max})$, capital $k(t_{max})$, consumption $c(t_{max})$ and utility $U$ for the interaction parameter $\varepsilon_k$ and the reference parameter $\varepsilon_\rho$. CV is the square root of the variance divided by the mean value and it measures inequality. The horizontal axis is $\varepsilon_k$, the first vertical axis is the mean and the second is the CV. The mean for $\varepsilon_\rho = 0$ is shown in navy and the CV in light navy and the mean for $\varepsilon_\rho = 0.1$ in green and the CV in light green. Similarly, Figures 6E–6H show the calculation results for the interaction parameter $\varepsilon_c$. Figures 6A–6D show that the capital interaction $\varepsilon_k$ works in the direction of lowering the discount rate $\rho(t_{max})$, which raises capital $k(t_{max})$, consumption $c(t_{max})$ and utility $U$. The values of CV are generally small, as is the effect of the social norm reference $\varepsilon_\rho$. Comparing the CV of $k(t_{max})$ and the CV of $c(t_{max})$ shows that the former is larger. In Figures 6E–6H, the consumption interaction $\varepsilon_c$ raises the discount rate $\rho(t_{max})$ more than the capital interaction $\varepsilon_k$, which in turn lowers capital $k(t_{max})$, consumption $c(t_{max})$ and utility $U$. Compared to $\varepsilon_k$, $\varepsilon_c$ has a generally more significant impact on CV. The impact on the CV of capital is larger than that of consumption and utility. The social norm illumination $\varepsilon_\rho$ reduces the increase in the discount rate $\rho(t_{max})$ and increases capital $k(t_{max})$, consumption $c(t_{max})$ and utility $U$. For $\varepsilon_c = 0.3$ and $\varepsilon_\rho = 0$, the CV of $k(t_{max})$ reaches 0.225 for the mean 1.82, corresponding to an increase in the Gini index of 0.14 when calculated from the distribution of $k(t_{max})$. When $\varepsilon_\rho = 0$ changes to 0.1, the CV drops to 0.09 *and* the incremental Gini index falls to 0.05. These results suggest that higher discount rates create capital (wealth) inequality and that social norms that discourage waste are essential.

Figure 7 plots capital $k(t_{max})$ and consumption $c(t_{max})$ for discount rate $\rho(t_{max})$. Figures 7A and 7B show the capital interactions $\varepsilon_k = 0.1$ and $\varepsilon_c = \varepsilon_\rho = 0$. Figures 7C and 7D show the case where the settings of 7A and 7B change to $\varepsilon_k = 0.3$. Figures 7E and 7F show the consumption interaction $\varepsilon_c = 0.1$ and $\varepsilon_k = \varepsilon_\rho = 0$. Figures 7G and 7H show the case where the settings of 7E and 7F are changed to $\varepsilon_c = 0.3$. The plots of $k(t_{max})$ in Figures 7A, 7C, 7E and 7G generally ride on the curve represented by equation (5). Furthermore, the plots of $c(t_{max})$ in Figures 7B, 7D, 7F and 7H generally ride on the curve represented by equation (6) (see graph of $k^*$ and $c^*$ at the saddle point for $\rho$ in Figure 11A). In Figures 7A–7D, the plots are scattered on the lower side of the curve and in Figures 7E–7H, the plots are scattered on the upper side of the curve. In the former, $\rho(t_{max})$ decreases with $\varepsilon_k$, showing the intermediate steps until convergence to a new saddle point, where capital $k$ and consumption $c$ become larger. In the latter, $\rho(t_{max})$ increases with $\varepsilon_c$, showing the intermediate steps until convergence to a new saddle point, where capital $k$ and consumption $c$ become smaller. Comparing Figures 7G and 7H with Figures 7C and 7D, the impact of $\varepsilon_c$ on the discount rate $\rho(t_{max})$ is more significant than that of $\varepsilon_k$. The curved shape is more clearly visible in Figure 7G and 7H than in Figure 7C and 7D because the range of the discount rate $\rho(t_{max})$ is wider due to $\varepsilon_c$ than due to $\varepsilon_k$.

Figure 8 plots utility $U$ over an infinite horizon against capital $k(t_{max})$ and consumption $c(t_{max})$. Figures 8A and 8B show the capital interactions $\varepsilon_k = 0.1$ and $\varepsilon_c = \varepsilon_\rho = 0$. Figures 8B and 8C show the case where the settings of 8A and 8B change to $\varepsilon_k = 0.3$ and Figures 8E and 8F show the consumption interaction $\varepsilon_c = 0.1$ and $\varepsilon_k = \varepsilon_\rho = 0$. Figures 8G and 8H show the case where the settings of 8E and 8F change to $\varepsilon_c = 0.3$. Comparing Figures 8C and 8D with Figures 8A and 8B shows that as the capital interaction $\varepsilon_k$ increases, utility $U$ increases overall and its distribution widens. Comparing Figures 8G and 8H with Figures 8E and 8F shows that as the consumption interaction $\varepsilon_c$ increases, utility $U$ decreases overall and the shape of its distribution changes. Comparing Figures 8A–8D and Figures 8E–8H reveals difference in the tendency of $\varepsilon_k$ and $\varepsilon_c$ to affect the distribution of utility $U$. For $\varepsilon_k$, the distribution moves toward the upper right, where $U$ and $k(t_{max})$ or $c(t_{max})$ become larger. In contrast, for $\varepsilon_c$, the distribution once moves toward the upper left, where $U$ increases and $k(t_{max})$ or $c(t_{max})$ decreases while $\varepsilon_c$ is small. It then moves toward the lower left, where $U$ and $k(t_{max})$ or $c(t_{max})$ decrease as $\varepsilon_c$ becomes large.

Figures 9A and 9B are plots of $\rho(t_{max})$ versus $k(t_{max})$ and $c(t_{max})$ versus $U$ for $\varepsilon_k = 0.3$ and $\varepsilon_c = \varepsilon_\rho = 0$, respectively. Figure 9A is a repeat of 7C and 9B is a repeat of 8D. Similarly, Figures 9C and 9D are for $\varepsilon_c = 0.3$ and $\varepsilon_k = \varepsilon_\rho = 0$ and are repeats of Figures 7G and 8H, respectively. Figures 9E and 9D show when the settings of either $\varepsilon_k = 0.3$ or $\varepsilon_c = 0.3$ are randomly chosen on the interaction occasion every period $t_p$. Figures 9G and 9H are the cases where the settings of Figures 9E and 9D change from $\varepsilon_\rho = 0$ to 0.1 and a reference to the social norm is added. The shape of Figure 9E is intermediate between 9A and 9C and the shape of Figure 9F is intermediate between 9B and 9D. The means of $\rho(t_{max})$, $k(t_{max})$, $c(t_{max})$ and $U$ in Figures 9E and 9F are intermediate between those values in Figures 9A–9D (roughly the average of the values shown in 9A–9D). Although not shown in Figure 9, the CVs are intermediate but larger than the average values in 9A–9D. Figures 9G and 9H change to $\varepsilon_\rho = 0.1$ and also have intermediate shapes and values like 9E and 9F; however, the reference to social norms results in a smaller mean for $\rho(t_{max})$ and larger means for $k(t_{max})$, $c(t_{max})$ and $U$ compared to 9E and 9F. Although not shown in Figure 9, all CVs for $\varepsilon_\rho = 0.1$ are smaller than that for $\varepsilon_\rho = 0$.

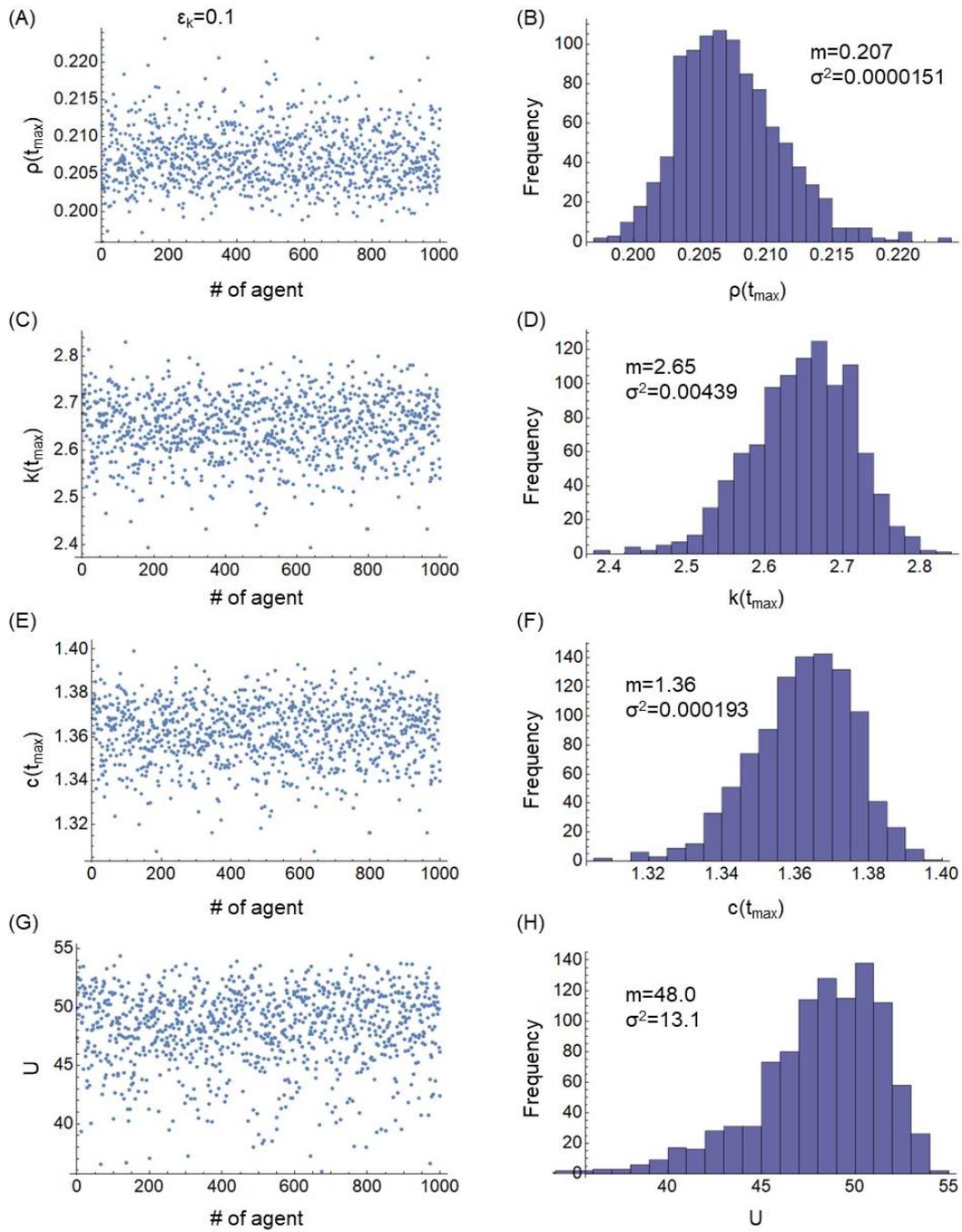

**Figure 2:** Calculation results for capital interactions

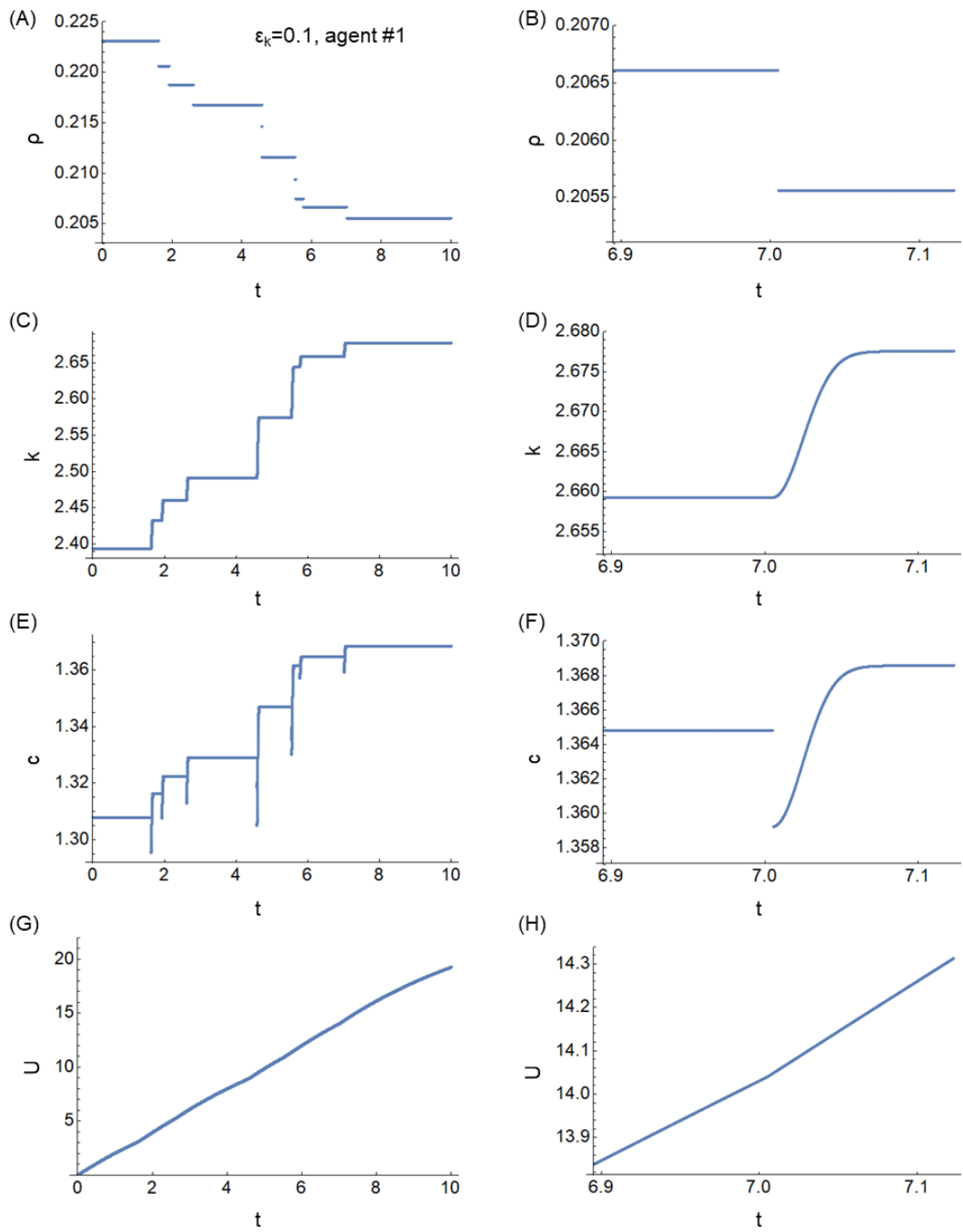

**Figure 3:** Temporal waveforms of capital interactions

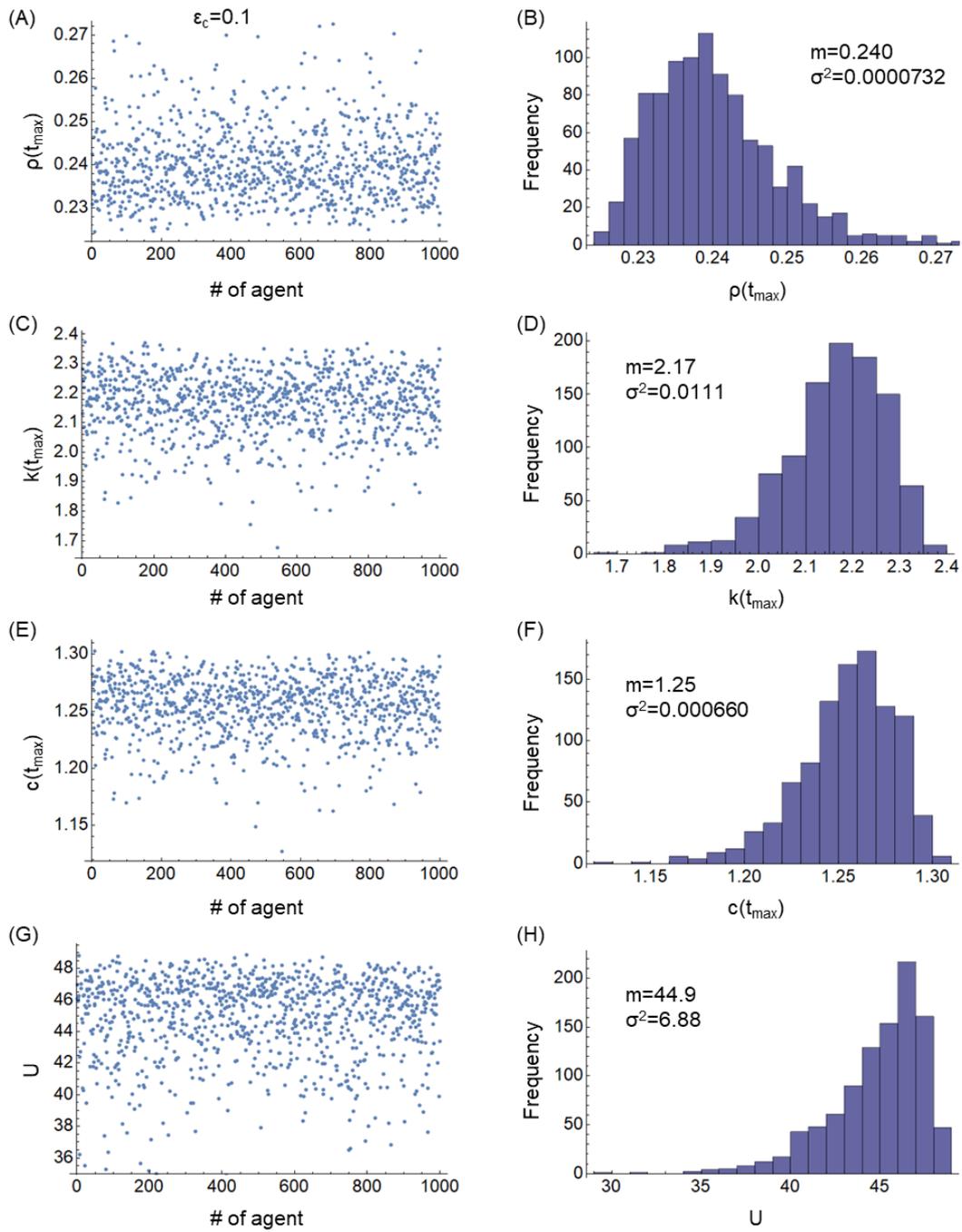

**Figure 4:** Calculation results for consumption interactions

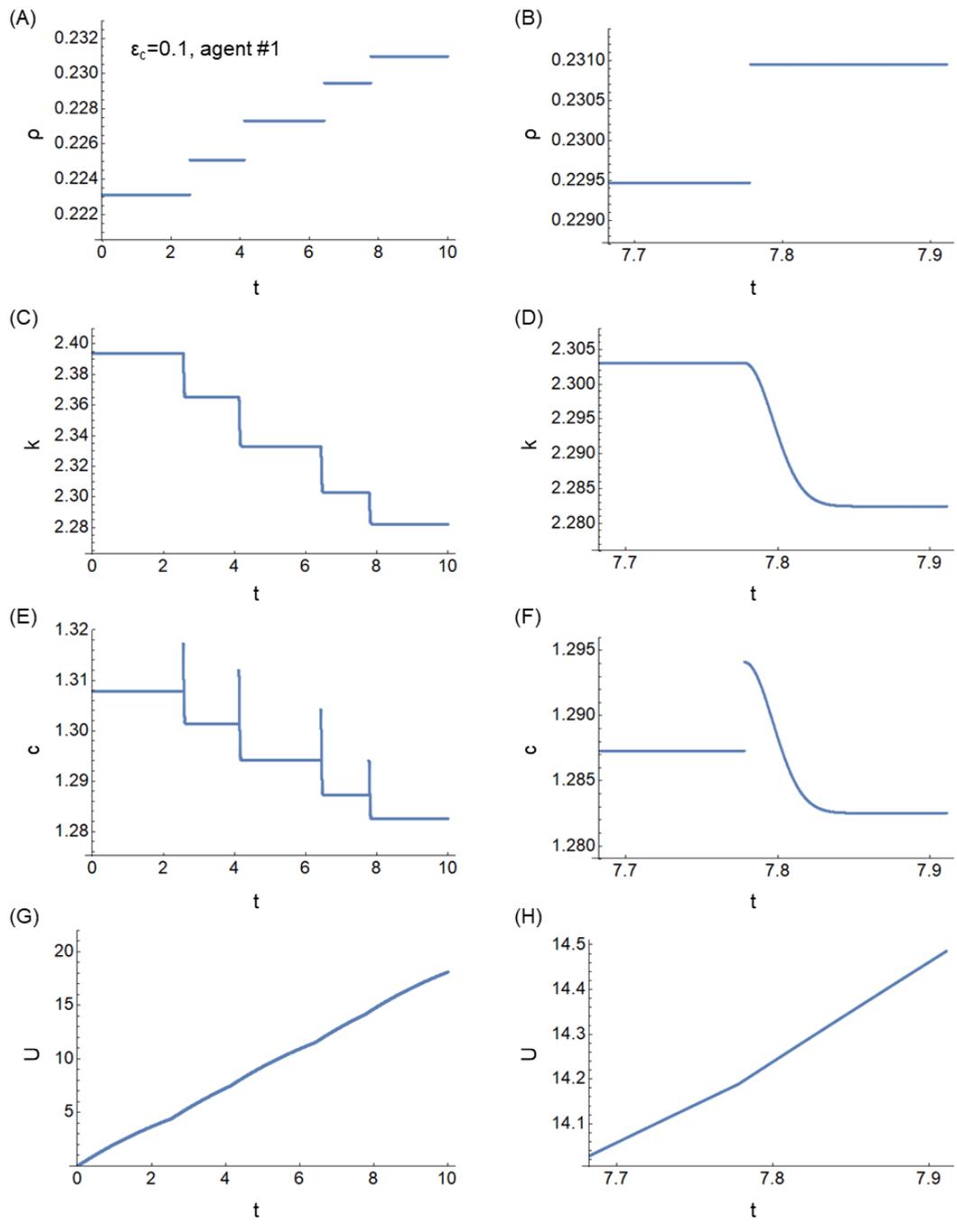

**Figure 5:** Temporal waveforms of consumption interactions

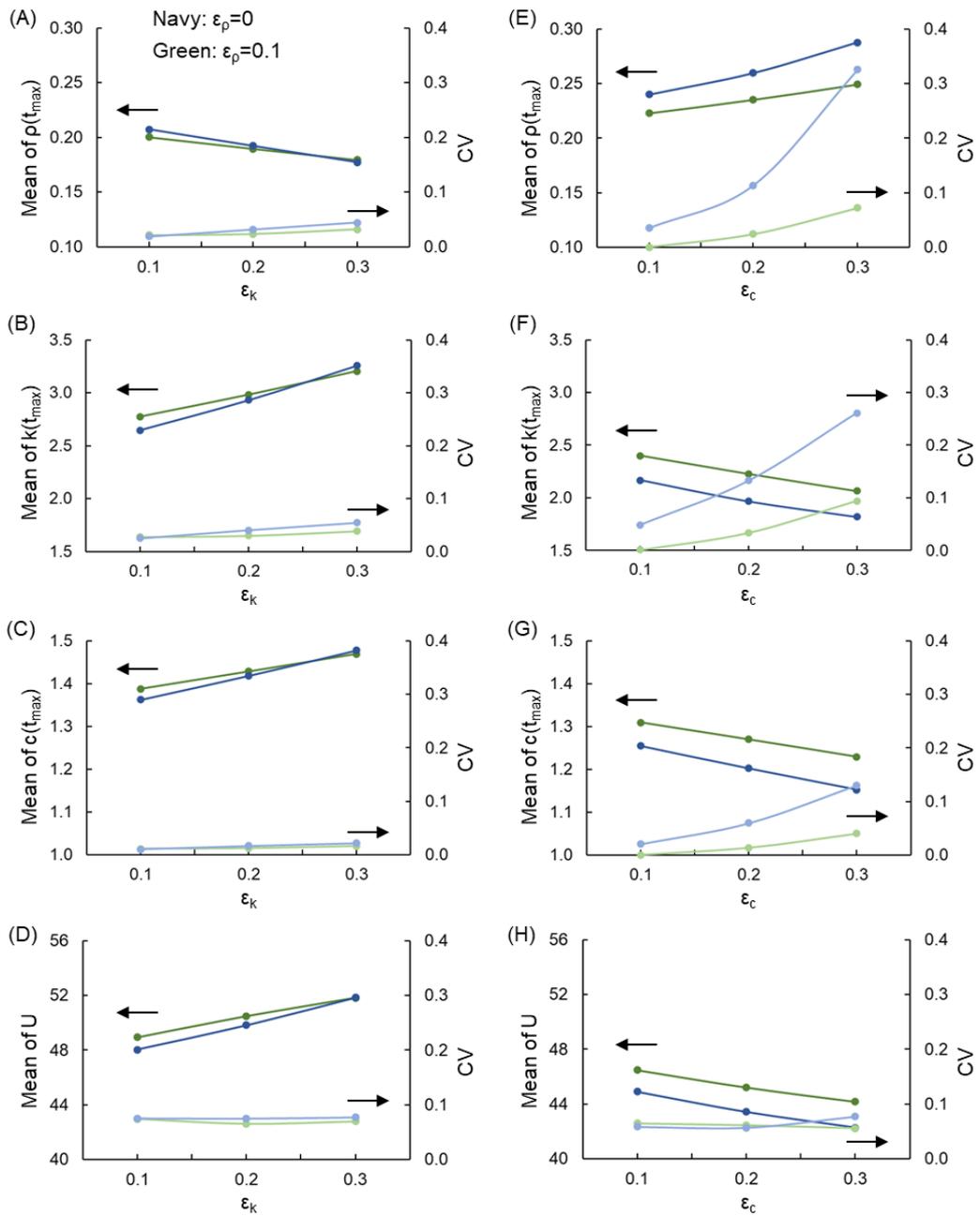

**Figure 6:** Mean and coefficient of variation (CV) of discount rate $\rho(t_{max})$, capital $k(t_{max})$, consumption $c(t_{max})$ and utility $U$ for others' interaction parameter $\varepsilon_k$, $\varepsilon_c$ and social norm reference parameter $\varepsilon_\rho$

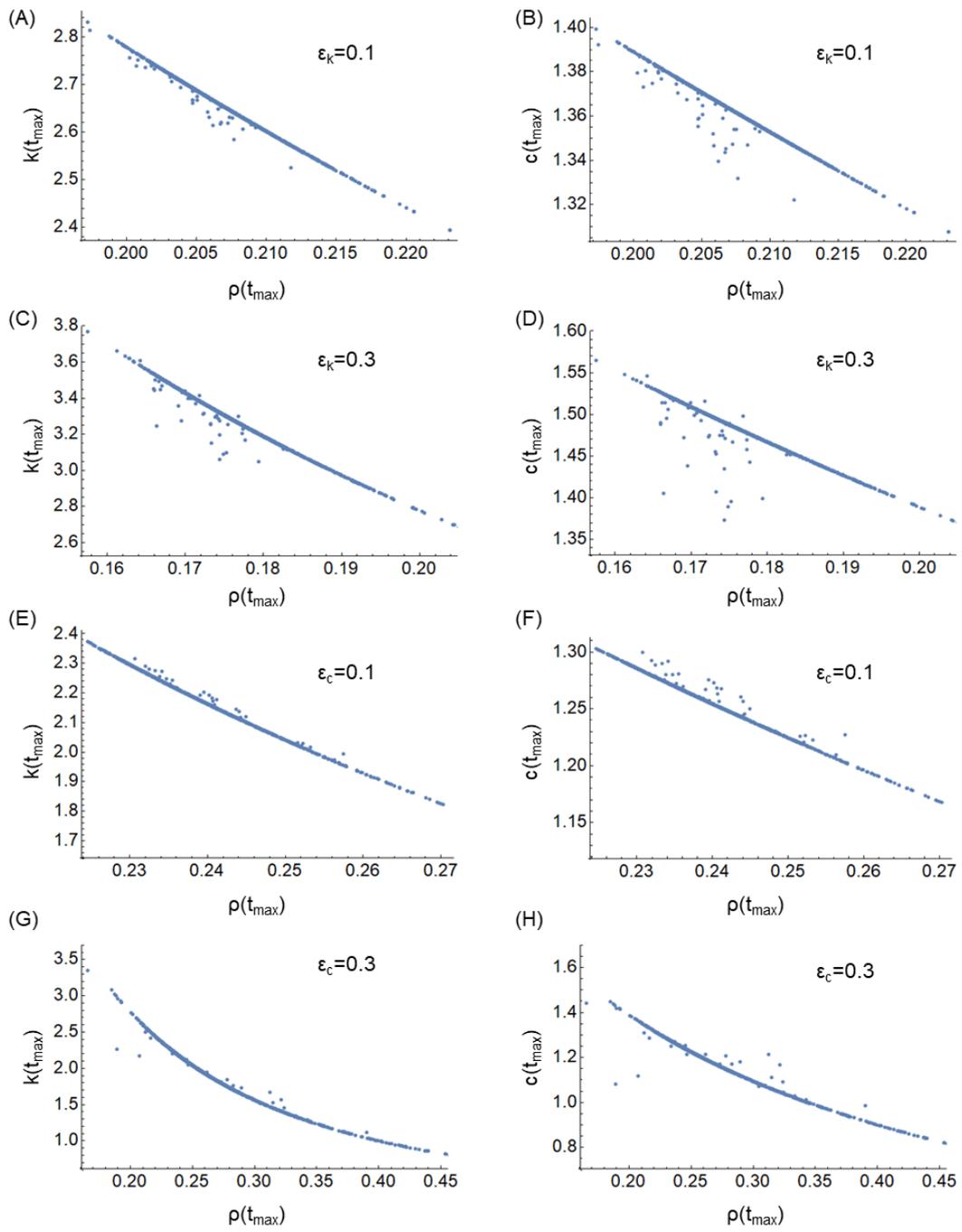

**Figure 7:** Capital $k(t_{max})$ and consumption $c(t_{max})$ for discount rate $\rho(t_{max})$

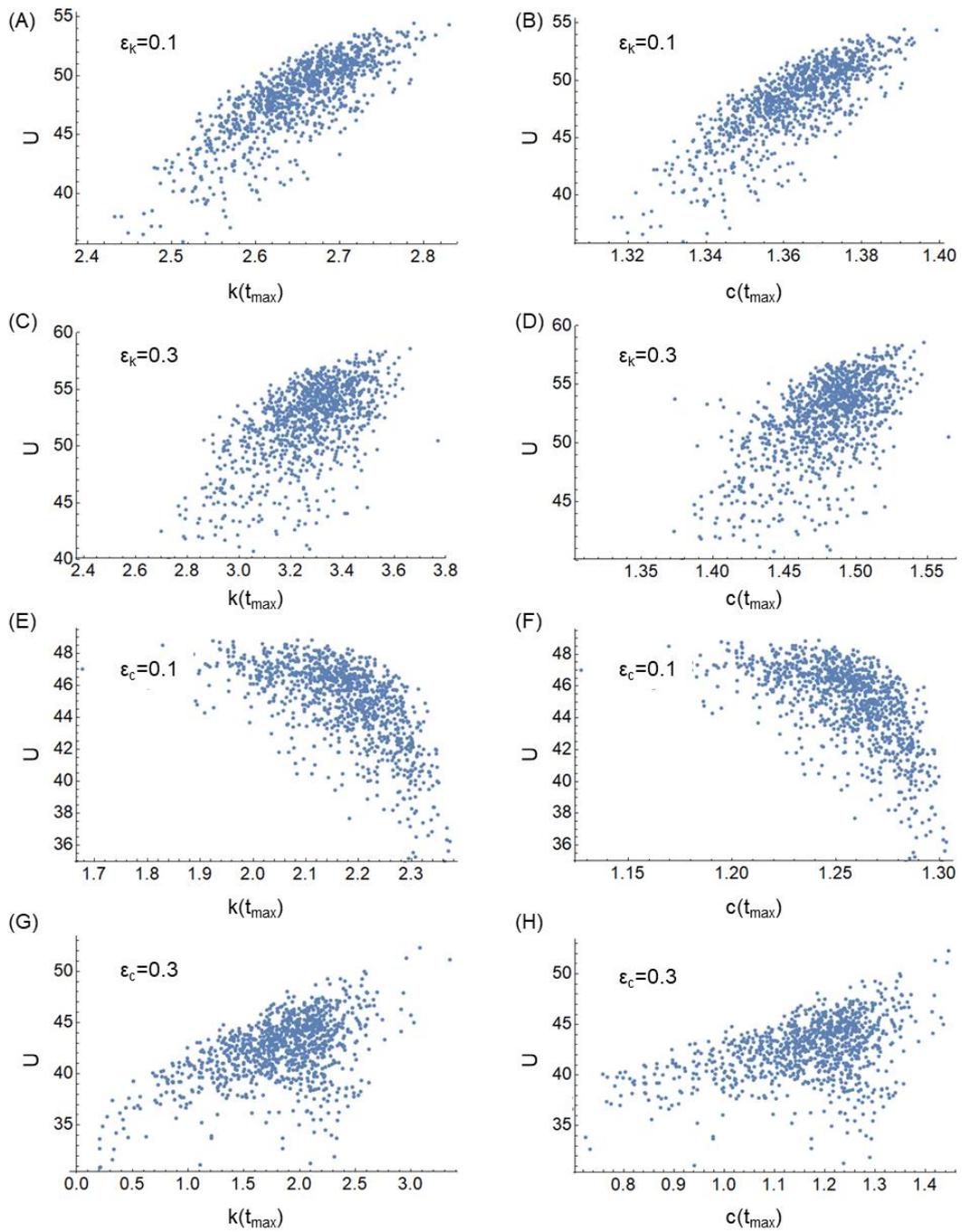

**Figure 8:** Utility $U$ for capital $k(t_{max})$ and consumption $c(t_{max})$

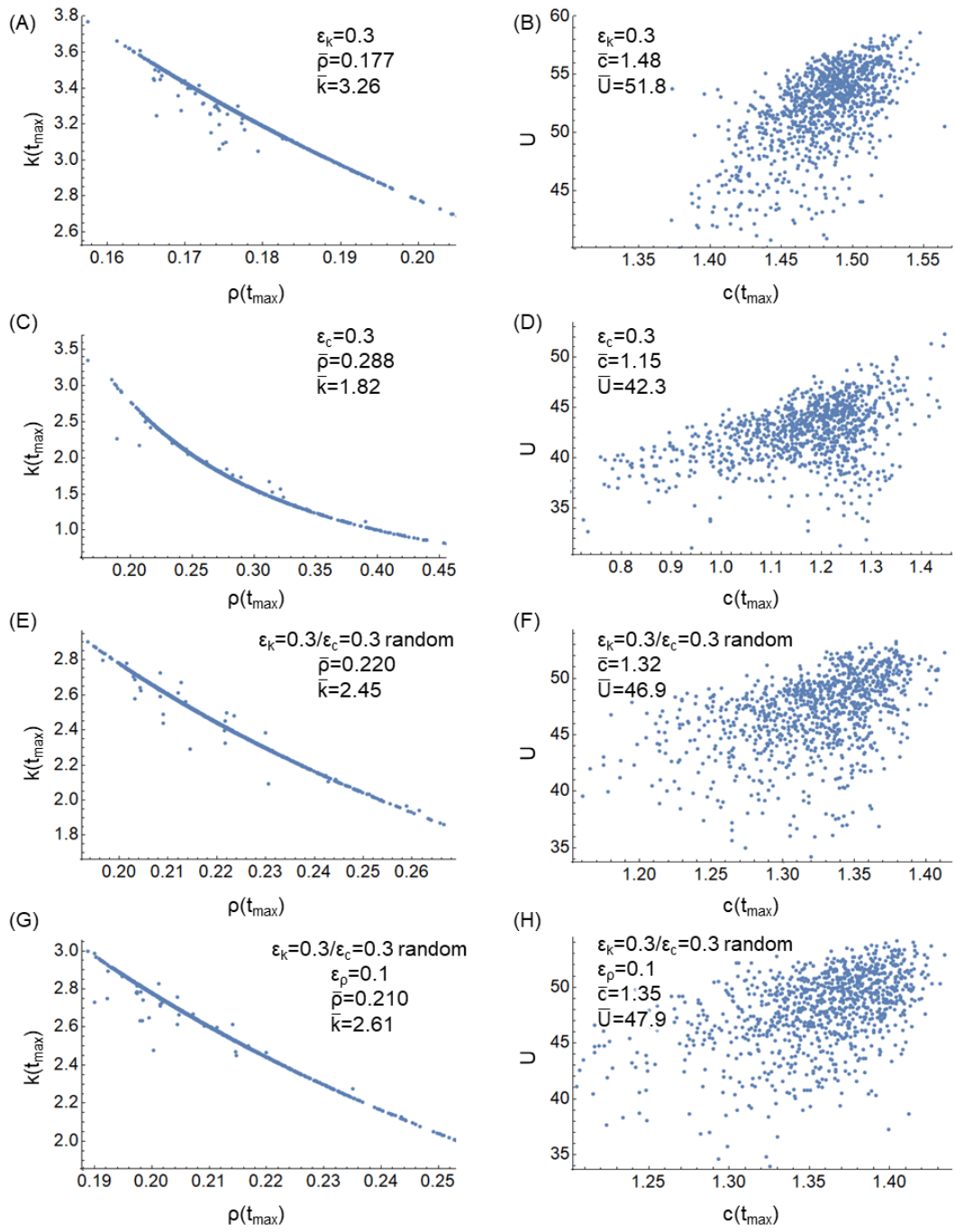

**Figure 9:** Discount rate $\rho(t_{max})$ versus capital $k(t_{max})$ and consumption $c(t_{max})$ versus utility $U$

# DISCUSSION

We formulated changes in the discount rate (time preference) based on the intercomparison of capital or consumption among agents and reference to social norms by connecting the Ramsey–Cass–Koopmans model of dynamic macroeconomics and the agent interaction model of microeconomics. This approach revealed that interactions with others affect the discount rate and produce inequality in capital (wealth) and utility and that references to social norms reduce changes in the discount rate and inequality. Specifically, when the interaction parameter with others $\varepsilon_c = 0.3$ in Figure 6, the coefficient of variation increases to $0.225$ and the Gini index increases by $0.14$ as the discount rate $\rho$ changes. In this case, when the social norm reference parameter $\varepsilon_\rho = 0.1$, the coefficient of variation drops to $0.09$ and the increment of the Gini index falls to $0.05$.

A cohort study (Epper *et al.*, 2020) divides participant groups into tertiles according to time preference (patience), showing that the wealth difference between the top and bottom time preference groups is in the 7–8 percentile. The difference between the top and bottom tertile is roughly 2σ (twice the standard deviation); thus, the wealth difference translates into a coefficient of variation of about 4%, which corresponds approximately to cases $\varepsilon_k \sim 0.3$ or $\varepsilon_c \sim 0.1$ in Figure 6. This result can be interpreted as follows: if every individual had the same time preference at birth, as they grow up, their time preferences would gradually diverge due to their interactions with others, resulting in wealth inequality.

A cross-cultural study on time preference (Reiger *et al.*, 2021) calculates universal time preference (UTP) for 117 countries by integrating various survey data, including discount rate data from reference (Wang *et al.* 2016). A larger UTP means a smaller discount rate. Figure 10 uses UTP data and World Bank Gini index data to UTP and the Gini index for 104 countries (The World Bank). UTP and the Gini index have a weak correlation, although considerable variation occurs among countries due to economic conditions and cultural differences. The regression line indicates that the difference in the Gini index relative to the maximum difference in UTP is about 0.15, which corresponds to the case $\varepsilon_c \sim 0.3$ in Figure 6. Social norms, including rules and culture, are fostered by mutual recognition (Heath, 2008); thus, we can infer that at least part of the difference in the Gini index is caused by differences in time preference due to interaction with others. The red dots in Figure 10 indicate Islamic states, which are roughly below the regression line. This finding may be because Islamic states have laws (*Sharia*) that discourage waste compared to capitalist states.

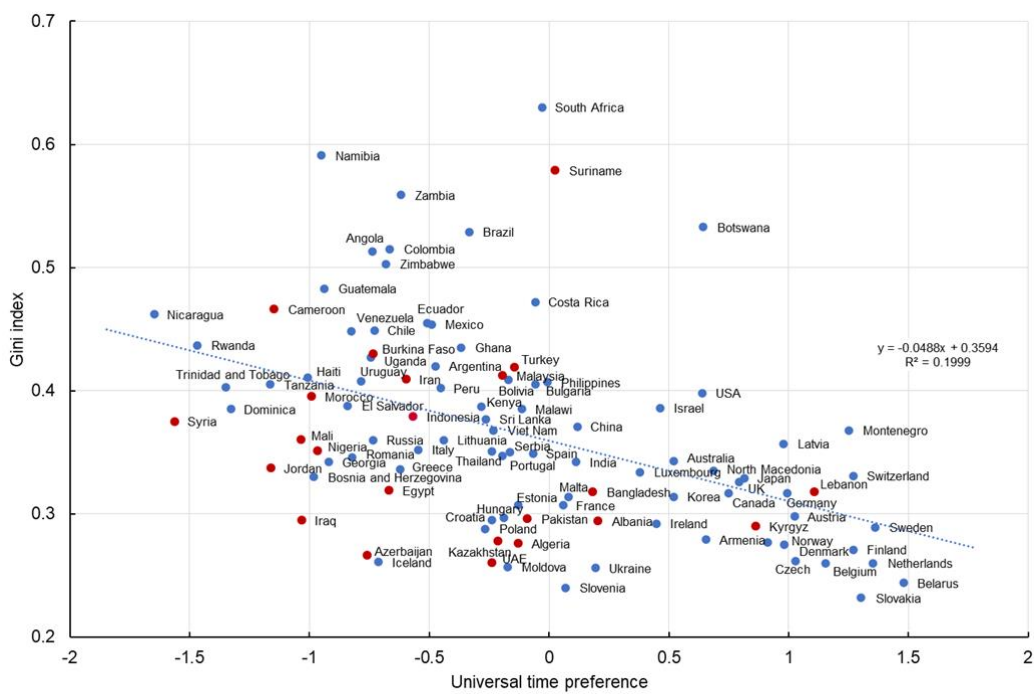

**Figure 10:** Universal time preference and Gini Index

Comparing Figures 6B and 6C and Figures 6F and 6G shows that the impact of the coefficient of variation of capital $k$ from its interaction with others, $\varepsilon_k$ and $\varepsilon_c$, is larger than that of consumption $c$. This finding is explained in Figure 11. Figure 11A graphs capital $k^*$ and consumption $c^*$ at the saddle point concerning the discount rate $\rho$ using equations (5) and (6); 11B is a graph depicting the derivative $k^{*\prime}$ of $k^*$ and the derivative $c^{*\prime}$ of $c^*$ for $\rho$. Figure 11B shows that the absolute value of $k^{*\prime}$ is larger than the absolute value of $c^{*\prime}$; therefore, the impact of $k$ from a change in $\rho$ is larger than that of $c$. This outcome suggests that interaction with others has a more significant impact on capital (wealth) inequality than consumption. The impact of the CV of utility $U$ from $\varepsilon_k$ and $\varepsilon_c$ is even smaller than that of $k$ and $c$ because the impact of $c$ is smaller than that of $k$, $U$ is a function of $c$ and the integral value $U$ from time $t = 0$ is less sensitive than the instantaneous value of $k$ and $c$. This result is reflected in the slight inflexion of the curve in Figures 3G and 5G. Figures 6D and 6H show that the CV of $U$ is about 8%, e.g. the CV of the well-being measured value in the literature (Pontin et al., 2013) is 23%. Among the various constituent factors of well-being, utility based on consumption accounts for about 1/3 of well-being; therefore, the value of the CV of $U$ itself is not small.

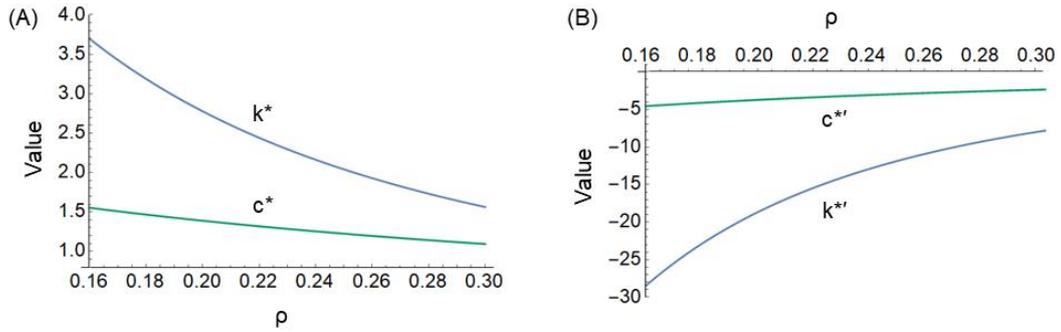

**Figure 11:** Capital $k^*$ and consumption $c^*$ at saddle point for discount rate $\rho$

Comparing the left and right pairs in Figure 6, such as 6A and 6E, 6B and 6F and 6C and 6G, the impact of $\varepsilon_c$ on the mean and coefficient of variation of $\rho(t_{max})$, $k(t_{max})$, $c(t_{max})$ and $U$ is larger than that of $\varepsilon_k$ because equations (20) and (21) represent interactions. Looking at these, $\varepsilon_c$ works in the direction of increasing the discount rate $\rho$ and $\varepsilon_k$ decreases it. $\rho_{NEW}$ is expressed as the multiplication of $\varepsilon_c$ and $\varepsilon_k$ concerning the original $\rho_{OLD}$; therefore, as the interaction is repeated, the difference between $\rho_{NEW}$ and $\rho_{OLD}$ ($\rho_{NEW} - \rho_{OLD}$) gradually increases for $\varepsilon_c$ and decreases for $\varepsilon_k$, i.e. the impact of $\varepsilon_c$ on the change in $\rho$ is larger than that of $\varepsilon_k$. This finding suggests that consumption interactions impact capital (wealth) inequality more than capital interactions.

Figure 6 shows the impacts of the social norm reference $\varepsilon_\rho$ on $\rho(t_{max})$, $k(t_{max})$, $c(t_{max})$ and $U$ are larger on the $\varepsilon_c$ side and smaller on the $\varepsilon_k$ side because, as mentioned in the above paragraph, the change in $\rho$ is larger for $\varepsilon_c$ than for $\varepsilon_k$. Therefore, the suppression effect of the change in $\rho$ by $\varepsilon_\rho$ is more likely to appear.

Furthermore, the slight change in $\rho$ due to $\varepsilon_k$ means that the value of $\rho$ is near the initial value $\rho_0 = 0.223$ and close to the normative discount rate $\rho_{NORM} = 0.2$ because this makes the $(\rho_{NORM} - \rho)/\rho$ term in equations (20) and (21) smaller. These findings suggest that social norms are more effective in reducing the impact of consumption interactions on inequality than capital interactions. Note that if the $\rho_{NORM}$ setting is too large (i.e. if the social norm is made too loose), $\rho$ will be larger and $k$ and $c$ will be smaller for $\varepsilon_\rho = 0.1$ than for $\varepsilon_\rho = 0$; this is shown in the case of $\varepsilon_k = 0.3$ in Figures 6A–6D. Conversely, a small setting of $\rho_{NORM}$ reduces waste, but an excessive setting may lead to stinginess. Therefore, the setting of $\rho_{NORM}$ should be set between stinginess and wastefulness. Greek philosophy, Buddhism and Islamic law also teach this.

The difference in the shape of the distribution of utility $U$ between Figures 8A–8D and Figures 8E–8H can be interpreted as follows. In Figures 8A–8D, the new saddle points, $k^*$ and $c^*$, are larger because the capital interaction $\varepsilon_k$ decreases the discount rate $\rho$ and utility $U$ increases accordingly, i.e. it has a vector in the upper right direction of the graph. Nonetheless, since a decrease in $\rho$ temporarily lowers consumption $c$, $U$, a function of $c$, goes down; therefore, it has a vector in the lower direction of the graph. Thus, the distribution of Figures 8A–8D will have superimposed vectors in the upper right and lower directions. Conversely, in Figures 8E–8H, the consumption interaction $\varepsilon_c$ increases $\rho$, $k^*$, $c^*$ and $U$ decrease and those distributions have vectors in the lower left direction of the graph; however, since an increase in $\rho$ temporarily raises consumption $c$, $U$, a function of $c$, increases and has a vector in the upper direction of the graph. Thus, the distribution of Figures 8E–8H will have superimposed vectors in the lower left and upper directions. Previous studies have examined income and well-being relationships (Killingsworth, 2023); in the future, it might be fruitful to relate this to the graph of capital $k$ and utility $U$ in Figure 8.

Figure 9E, with randomly selected capital interaction $\varepsilon_k$ and consumption interaction $\varepsilon_c$, has a distribution shape intermediate between Figures 9A and 9C and a roughly mean intermediate value between them. Figure 9F also has a shape intermediate between Figures 9B and 9D and a roughly mean intermediate value between them. These indicate that the impacts of $\varepsilon_k$ and $\varepsilon_c$ on the discount rate $\rho$, capital $k$,

consumption $c$ and utility $U$ can be treated as linear independence. The linearity of the graph of means in Figure 6 also indicates that the impacts of $\varepsilon_k$ and $\varepsilon_c$ on $\rho$, $k$, $c$ and $U$ can be extrapolated. These are useful in examining various combinations of $\varepsilon_k$ and $\varepsilon_c$. If we can understand whether individuals value capital or consumption in their interactions with others, we can approximate the impact on capital (wealth) and consumption. Although the coefficient of variation is an intermediate value, it is not an average value and cannot be treated as linear independence; therefore, it must be recalculated again when examining capital (wealth) and utility inequality.

The distribution shapes in Figures 2 and 4 show that the capital interaction $\varepsilon_k$ and the consumption interaction $\varepsilon_c$ generate distributions, i.e. inequalities, from initial uniform values; however, they are not power distributions, e.g. Pareto distribution (Newman, 2005), commonly found in wealth and income distributions. In other words, it is reasonable to consider that inequalities brought by changes in time preference (discount rate) based on interactions with others are part of the overall inequality. In contrast, econophysics studies various distributions, such as power and gamma distributions, by modelling asset exchange among agents based on the analogy of kinetic energy exchange (Chatterjee & Chakrabarti, 2007; Chakrabarti, AS & BK 2010); however, since these studies are physical analogies, they do not provide the same insight into capital, consumption and utility dynamics as macroeconomics. To study inequality due to factors other than time preference and macroeconomic dynamics, developing an approach that connects the microeconomic interactions with dynamic macroeconomic models would be helpful, as presented in this study. This approach would lead to generalizations beyond the findings of this study. Based on this study, instead of the interaction model of time preference, it is possible to model the exchange of products (market) related to capital (Chatterjee & Chakrabarti, 2007; Chakrabarti AS & BK, 2010) and to model the interest and debt on capital, the production by joint ventures that combine capital, the redistribution and mutual aid of capital (Kato & Hiroi, 2021; Kato, 2022; Kato, 2023). As a future development potential, this study's approach could be further extended by incorporating into the utility function not only consumption but also natural capital, social common capital and social relationship capital related to well-being.

# CONCLUSION

This study presented a new model that connects the interaction effects of intercomparisons between individuals and others and references to social norms on time preferences to a dynamic macroeconomic model. Through this, we estimated the impact of changes in time preference (discount rate) on capital, consumption and utility and the inequality (coefficient of variation, Gini index) resulting from these changes. The new findings are that (i) interactions with others significantly impact capital (wealth) inequality more than consumption. (ii) Among interactions with others, intercomparisons of consumption have a more significant impact on capital inequality and (iii) social norms effectively reduce the impact of consumption on inequality. These findings reaffirm the importance of social norms against waste. Our supporting evidence indicates that the quantitative impact of the time preference interaction on wealth inequality corresponds to survey data from cohort and cross-cultural studies and that it has no small impact on utility inequality.

Previous studies have shown that multiple psychological motivations and factors, as well as mutual recognition and shared expectations of social groups, are essential for time preference; however, no specific mathematical model exists for these factors and time preference. Therefore, we formulated the interactions on time preferences as relative capital, consumption and social norm comparisons. Future behavioural economic research on time preferences could improve this study's model to provide further insights. Moreover, this study's approach of connecting microeconomic interactions with dynamic macroeconomic models can be expanded to model exchange, interest and debt, redistribution and mutual aid.

## ACKNOWLEDGMENTS

I thank Dr. Yosuke Tanabe, Chief Researcher, Center for Exploratory Research, Hitachi, Ltd., for discussions on macroeconomic and microeconomic models and Mr. Mohammad Rezoanul Hoque, PhD Student, Department of Economics, Texas Tech University, for discussions on reflecting social norms in macroeconomic models. Furthermore, I would also like to thank my colleagues at the Hitachi Kyoto University Laboratory of the Kyoto University Open Innovation Institute for their ongoing cooperation.

## STUDY FUNDING

No funding was received for conducting this study.

## APC FUNDING

No funding was received for APC.

## CONFLICT OF INTEREST

The author declares no conflict of interest associated with this manuscript.

## AUTHORS' CONTRIBUTIONS

Takeshi Kato (Conceptualization, Data curation, Formal Analysis, Investigation, Methodology, Software, Validation, Visualization, Writing—original draft, Writing—review & editing)

## DATA AVAILABILITY

The author will share the data underlying this article upon reasonable request.

# REFERENCES

Angle, J. (1986) 'The surplus theory of social stratification and the size distribution of personal wealth', *Social Forces*, 65/2: 293–326.

Aoki, M. (2001) *Toward a Comparative Institutional analysis*. Cambridge: MIT Press.

Becker, R. A. (1980) 'On the long-run steady state in a simple dynamic model of equilibrium with heterogeneous households', *The Quarterly Journal of Economics*, 95/2: 375–82. https://doi.org/10.2307/1885506

Boorman, E. D., Behrens, T. E. J., Woolrich, M. W., and Rushworth, M. F. S. (2009) 'How green is the grass on the other side? Frontopolar cortex and the evidence in favor of alternative courses of action', *Neuron*, 62/5: 733–43. https://doi.org/10.1016/j.neuron.2009.05.014

Caiani, A., Godin, A., Caverzasi, E., Gallegati, M., Kinsella, S., and Stiglitz, J. E. (2016) 'Agent based-stock flow consistent macroeconomics: Towards a benchmark model', *Journal of Economic Dynamics and Control,* 69: 375–408. https://doi.org/10.1016/j.jedc.2016.06.001

Carroll, C. D., Overland, J., and Weil, D. N. (1997) 'Comparison utility in a growth model', *Journal of Economic Growth*, 2: 339–67. https://doi.org/10.1023/A:1009740920294

Cass, D. (1965) 'Optimum growth in an aggregative model of capital accumulation', *The Review of Economic Studies*, 32/3: 233–40. https://doi.org/10.2307/2295827

Chakrabarti, A. S., and Chakrabarti, B. K. (2009) 'Microeconomics of the ideal gas like market models', *Physica A*, 388/19: 4151–8. https://doi.org/10.1016/j.physa.2009.06.038

Chakrabarti, A. S., and Chakrabarti, B. K. (2010) 'Statistical theories of income and wealth distribution', *Economics*, 4: 1–32. https://doi.org/10.5018/economics-ejournal.ja.2010-4

Chakraborti, A. (2002) 'Distributions of money in model markets of economy', *International Journal of Modern Physics C*, 13: 1315–21.

https://doi.org/10.1142/S0129183102003905

Chancel, L., Piketty, T., Saez, E., and Zucman, G., eds. World Inequality Lab wir2022.wid.world. (2022) *World Inequality Report 2022* [Internet]. https://wir2022.wid.world/www-site/uploads/2022/03/0098-21_WIL_RIM_RAPPORT_A4.pdf

Chang, W-Y., Tsai, H-F., and Chang, J-J. (2011) 'Interest-rate rules, and indeterminacy', *The Japanese Economic Review*, 62: 348–64. https://doi.org/10.1111/j.1468-5876.2011.00540.x

Chatterjee, A., and Chakrabarti, B. K. (2007) 'Kinetic exchange models for income and wealth distributions', *The European Physical Journal B*, 60: 135–49. https://doi.org/10.1140/epjb/e2007-00343-8

Cohen, J., Ericson, K. M., Laibson, D., and White, J. M. (2020) 'Measuring Time Preferences', *Journal of Economic Literature*, 58/2: 299–347. https://doi.org/10.1257/jel.20191074

De Grauwe, P. (2010) 'Top-down versus bottom-up macroeconomics', *CESifo Working Paper*, 3020. https://www.cesifo.org/en/publications/2010/working-paper/top-down-versus-bottom-macroeconomics

Dunbar, R. I. M. (2003) 'The Social brain: Mind, language, and society in evolutionary perspective', *Annual Review of Anthropology*, 32: 163–81. https://doi.org/10.1146/annurev.anthro.32.061002.093158

Elminejada, A., Havraneka, T., and Irsovaa, Z. (2022) 'Relative risk aversion: A meta-analysis', *EconStor*, 260586 [Preprint]. http://hdl.handle.net/10419/260586

Epper, T., Fehr, E., Fehr-Duda, H., Kreiner, C. T., Lassen, D. D., Leth-Petersen, S., and Rasmussen, G. N. (2020) 'Time discounting and wealth inequality', *American Economic Review*, 110/4: 1177–205. https://doi.org/10.1257/aer.20181096

Epstein, L. G. (1987) 'A simple dynamic general equilibrium model', *Journal of Economic Theory*, 41/1: 68–95. https://doi.org/10.1016/0022-0531(87)90006-8

Frederick, S., Loewenstein, G., and O'Donoghue, T. (2002) 'Time discounting and time preference: A critical review', *Journal of Economic Literature*, 40/2: 351–401. http://www.jstor.org/stable/2698382

Guala, S. (2009) 'Taxes in a wealth distribution model by inelastically scattering of particles', *Interdisciplinary Description of Complex Systems*, 7/1, 1–7. https://www.indecs.eu/2009/indecs2009-pp1-7.pdf

Gualdi, S., Tarzia, M., Zamponi, F., and Bouchaud, J-P. (2015) 'Tipping points in macroeconomic agent-based models', *Journal of Economic Dynamics and Control*, 50: 29–61. https://doi.org/10.1016/j.jedc.2014.08.003

Haldane, A. G., and Turrell, A. E. (2018) 'An interdisciplinary model for macroeconomics', *Oxford Review of Economic Policy*, 34/1-2: 219–51. https://doi.org/10.1093/oxrep/grx051

Hartog, J., Ferrer-i-Carbonell, A., and Jonker, N. (2002) 'Linking measured risk aversion to individual characteristics', *KYKLOS*, 55/1: 3–26. https://doi.org/10.1111/1467-6435.00175

Hasumi, R. (2020) *Introduction to dynamic macroeconomics*. Japanese ed. Tokyo: Nippon Hyoron Sha.

Heath, J. (2008) *Following the rules: Practical reasoning and deontic constraint*. Oxford: Oxford University Press.

Heathcote, J., Storesletten, K., and Violante, G. L. (2009) 'Quantitative macroeconomics with heterogeneous households', *Annual Review of Economics*, 1/1: 319–54. https://doi.org/10.1146/annurev.economics.050708.142922

Hendricks, L. (2007) 'How important is discount rate heterogeneity for wealth inequality?', *Journal of Economic Dynamics and Control*, 31/9: 3042–68. https://doi.org/10.1016/j.jedc.2006.10.002

Itou, S. (2008) 'Research on relative risk aversion for relevance judgment', *Financial Planning Reaserch*, 8: 4–21. https://www.jasfp.jp/img/file21.pdf

Kato, T., and Hiroi, Y. (2021) 'Wealth disparities and economic flow: Assessment using

an asset exchange model with the surplus stock of the wealthy', *PLoS ONE*, 16/11: e0259323. https://doi.org/10.1371/journal.pone.0259323

Kato, T. (2022) 'Islamic and capitalist economies: Comparison using econophysics models of wealth exchange and redistribution', *PLoS ONE*, 17/9: e0275113. https://doi.org/10.1371/journal.pone.0275113

Kato, T. (2023) 'Wealth redistribution and mutual aid: Comparison using equivalent/non-equivalent exchange models of econophysics', *Entropy*, 25/2: 224. https://doi.org/10.3390/e25020224

Killingsworth, M. A., Kahneman, D., and Mellers, B. (2023) 'Income and emotional well-being: A conflict resolved', *PNAS*, 120/10: e2208661120. https://doi.org/10.1073/pnas.2208661120

Koopmans, T. C. (1965) 'On the concept of optimal economic growth', in Johansen, J. ed. *The econometric approach to development planning*, pp. 225–287. Amsterdam: North-Holland Publishing Company.

Krusell, P., and Smith, A. A., Jr. (1998) 'Income and wealth heterogeneity in the macroeconomy', *Journal of Political Economy*, 106: 867–96. https://doi.org/10.1086/250034

Laibson, D. (1997) 'Golden eggs and hyperbolic discounting', *The Quarterly Journal of Economics*, 112/2: 443–78. http://dx.doi.org/10.1162/003355397555253

Lawrance, E. C. (1991) 'Poverty and the rate of time preference: Evidence from panel data', *Journal of Political Economy*, 99/1: 54–77. https://doi.org/10.1086/261740

Markus, H., and Wurf, E. (1987) 'The dynamic self-concept: A social psychological perspective', *Annual Review of Psychology*, 38: 299–337. https://doi.org/10.1146/annurev.ps.38.020187.001503

Nakata, M. (2011) *Mathematics for dynamic macroeconomics from the basics*. Japanese ed. Tokyo: Nippon Hyoron Sha.

Newman, M. E. J. (2005) 'Power laws, pareto distributions and Zipf's law', *Contemporary Physics*, 46/5: 323–51. https://doi.org/10.1080/00107510500052444

Noritake, A., Ninomiya, T., and Isoda, M. (2018) 'Social reward monitoring and valuation in the macaque brain', *Nature Neuroscience*, 21: 1452–1462. https://doi.org/10.1038/s41593-018-0229-7

Obstfeld, M. (1990) 'Intertemporal dependence, impatience, and dynamics', *Journal of Monetary Economics*, 26/1: 45–75. https://doi.org/10.1016/0304-3932(90)90031-X

Pontin, E., Schwannauer, M., Tai, S., and Kinderman, P. (2013) 'A UK validation of a general measure of subjective well-being: The modified BBC subjective well-being scale (BBC-SWB)', *Health and Quality of Life Outcomes*, 11: 150. https://doi.org/10.1186/1477-7525-11-150

Prelec, D., and Loewenstein, G. (1991) 'Decision making over time and under uncertainty: A common approach', *Management Science*, 37/7: 770–86. https://doi.org/10.1287/mnsc.37.7.770

Ramsey, F. P. (1928) 'A mathematical theory of saving', *The Economic Journal*, 38: 543–559. http://dx.doi.org/10.2307/2224098

Rieger, M. O., Wang, M., and Hens, T. (2021) 'Universal time preference', *PLoS ONE*, 16/2: e0245692. https://doi.org/10.1371/journal.pone.0245692

Romer, D. (2019) *Advanced macroeconomics*. 5th ed. New York: McGraw-Hill.

Ryder, H. E., Jr. and Heal, G. M. (1973) 'Optimal growth with intertemporally dependent preferences', *The Review of Economic Studies*, 40/1: 1–31. https://doi.org/10.2307/2296736

Sedik, T. S., and Xu, R. (2020) 'A vicious cycle: How pandemics lead to economic despair and social unrest', *IMF Working Papers*, 2020/216. https://doi.org/10.5089/9781513559162.001

Samuelson, P. A. (1937) 'A note on measurement of utility', *The Review of Economic Studies*, 4/2: 155–161. https://doi.org/10.2307/2967612

Shi, S., and Epstein, L. G. (1993) 'Habits and time preference', *International Economic Review*, 34(1): 61–84. https://doi.org/10.2307/2526950

Strulik, H. (2008) 'Comparing consumption: A curse or a blessing?', *Leibniz Universitat Hannover Discussion Paper*, 382. http://dx.doi.org/10.2139/ssrn.1084882

Suen, R. M. H. (2014) 'Time preference and the distributions of wealth and income', *Economic Inquiry*, 52/1: 364–81. https://doi.org/10.1111/j.1465-7295.2012.00509.x

The World Bank. *Gini index* [Internet]. https://data.worldbank.org/indicator/SI.POV.GINI

Turrell, A. (2016) 'Agent-based models: understanding the economy from the bottom up', *Bank of England Quarterly Bulletin*, 56/4: 173–88. https://www.bankofengland.co.uk/quarterly-bulletin/2016/q4/agent-based-models-understanding-the-economy-from-the-bottom-up

United Nations Human Settlements Programme (2008) *State of the world's cities 2008/2009 — Harmonious cities*. Earthscan Publications: London, UK. https://unhabitat.org/state-of-the-worlds-cities-20082009-harmonious-cities-2

Uzawa, H. (1968) 'Time preference, the consumption function, and optimum asset holdings', in Wolfe, J. N. ed. *Value, capital and growth: Papers in honor of Sir John Hicks*, pp. 485-504. Edinburgh: Edinburgh University Press.

Wang, M., and Rieger, M. O., and Hens, T. (2016) 'How time preferences differ: Evidence from 53 countries', *Journal of Economic Psychology*, 52: 115–135. https://doi.org/10.1016/j.joep.2015.12.001

World Economic Forum. *8 men have the same wealth as 3.6 billion of the world's poorest people. We must rebalance this unjust economy* [Internet]. https://www.weforum.org/agenda/2017/01/eight-men-have-the-same-wealth-as-3-6-billion-of-the-worlds-poorest-people-we-must-rebalance-this-unjust-economy/

Zhang, W-B. (2012) 'Habits, saving propensity, and economic growth', *Scientific Bulletin – Economic Sciences*, 11/2: 3–15. http://economic.upit.ro/repec/pdf/2012_2_1.pdf